\documentclass[conference]{IEEEtran}
\IEEEoverridecommandlockouts

\usepackage{amssymb}
\usepackage[cmex10]{amsmath}
\usepackage{stfloats}
\usepackage{graphicx}
\usepackage{subfigure}
\usepackage{tabularx}
\usepackage{epsfig,epsf,color,balance,cite}
\usepackage{verbatim}
\usepackage{url}
\usepackage{bm}
\usepackage{booktabs}

\usepackage{algorithm}
\usepackage{algorithmic}

\usepackage{color}
\definecolor{myc1}{rgb}{0,0,0}

\begin{document}

\title{ 
{Semantic Communication with Probability Graph: A Joint Communication and Computation Design}
}

\author{
\IEEEauthorblockN{Zhouxiang Zhao\IEEEauthorrefmark{1},
                  Zhaohui Yang\IEEEauthorrefmark{1}\IEEEauthorrefmark{2},
                 Quoc-Viet Pham\IEEEauthorrefmark{3},
                 Qianqian Yang\IEEEauthorrefmark{1}\IEEEauthorrefmark{2},
                  and Zhaoyang Zhang\IEEEauthorrefmark{1}\IEEEauthorrefmark{2}
                 }
	\IEEEauthorblockA{
			$\IEEEauthorrefmark{1}$College of Information Science and Electronic Engineering, Zhejiang University, Hangzhou, China\\
			$\IEEEauthorrefmark{2}$Zhejiang Provincial Key Laboratory of Info. Proc., Commun. \& Netw. (IPCAN), Hangzhou, China\\
   $\IEEEauthorrefmark{3}$School of Computer Science and Statistics, Trinity College Dublin, Dublin 2, D02 PN40 Ireland\\
			E-mails: 
   zhouxiangzhao@zju.edu.cn,
   yang\_zhaohui@zju.edu.cn,
   viet.pham@tcd.ie,\\
   qianqianyang20@zju.edu.cn,
   ning\_ming@zju.edu.cn
		}
\thanks{This work is supported by Zhejiang Lab Program under grant K2023QA0AL02, and Zhejiang Science and Technology Program under grant 2023C01021.}
\vspace{-3em}
}

\maketitle

\begin{abstract}
In this paper, we present a probability graph-based semantic information compression system for scenarios where the base station (BS) and the user share common background knowledge. We employ probability graphs to represent the shared knowledge between the communicating parties. During the transmission of specific text data, the BS first extracts semantic information from the text, which is represented by a knowledge graph. Subsequently, the BS omits certain relational information based on the shared probability graph to reduce the data size. Upon receiving the compressed semantic data, the user can automatically restore missing information using the shared probability graph and predefined rules. This approach brings additional computational resource consumption while effectively reducing communication resource consumption. Considering the limitations of wireless resources, we address the problem of joint communication and computation resource allocation design, aiming at minimizing the total communication and computation energy consumption of the network while adhering to latency, transmit power, and semantic constraints. Simulation results demonstrate the effectiveness of the proposed system.
\end{abstract}

\IEEEpeerreviewmaketitle

\section{Introduction}
Over the past few decades, the development of mobile communication technologies has greatly contributed to the progress of human society. In the 1940s, Shannon proposed the information theory, which focused on quantifying the maximum data transmission rate that a communication channel could support. Guided by this seminal work, existing communication systems have been designed based on metrics which concentrate on transmission rate\cite{larsson2014massive}. With the rapid increase in demand for intelligent applications of wireless communication, the future communication network will change from a traditional architecture that simply pursues high transmission rate to a new architecture that is oriented to complete tasks efficiently\cite{2023big}. There will be more and more tasks that require low latency and high efficiency in future mobile information networks, posing a huge challenge \cite{gunduz2022beyond} to the existing mobile communication systems. This trend gives rise to a new communication paradigm, called \emph{semantic communication}. Semantic communication is expected to be a key technology for the future 6G mobile communication systems \cite{10024766,chaccour2022less}.

This paper utilizes knowledge graphs to represent the semantic information of text data. As a structured form of knowledge with high information density, knowledge graph is an important technical tool for extracting semantic information. A knowledge graph \cite{9416312} is a structured representation of facts, which consists of entities and relations. Entities can be real-world objects and abstract concepts, and relations represent concrete relationships between entities. Knowledge in the knowledge graph can be represented by triples in form of (\emph{head, relation, tail}), such as (\emph{banana, a kind of, fruit}).

Recently, several works studied a number of problems related to semantic communication. The authors in \cite{xie2021deep} proposed a deep learning (DL) based semantic communication system for text transmission. To measure the performance of semantic communication, the authors also designed a new metric called sentence similarity. In \cite{weng2021semantic}, a DL-enabled semantic communication system for speech signals was designed. In order to improve the recovery accuracy of speech signals, especially for the essential information, it was developed based on an attention mechanism by utilizing a squeeze-and-excitation network. The work in \cite{tong2021federated} enabled the transmission of audio semantic information which captures the contextual features of audio signals. To extract the semantic information from audio signals, a wave to vector architecture based autoencoder that consists of convolutional neural networks was proposed. However, these existing works in \cite{xie2021deep,weng2021semantic,tong2021federated} did not consider the computation energy consumption for the semantic communication system.

The main contribution of this paper is a novel text information extraction system based on probability graph, which jointly considers the communication and computation energy consumption of the semantic communication network. The key contributions are listed as follows:

\begin{itemize}
\item We propose a knowledge graph compression method based on the probability graph shared by the BS and the user. The probability graph is developed based on the knowledge graph, which expands the conventional triple to a quadruple by introducing the dimension of relation probability. With the shared probability graph, certain information in the knowledge graph corresponding to the text can be omitted, reducing the amount of data to be transmitted.
\item The problem of joint communication and computation resource allocation design of the semantic communication system is investigated. We formulate an optimization problem whose goal is to minimize the total communication and computation energy consumption of the network while adhering to latency, transmit power, and semantic constraints.
\end{itemize}


\section{System Model}
\begin{figure}[t]
\centering
\includegraphics[width=3.4in]{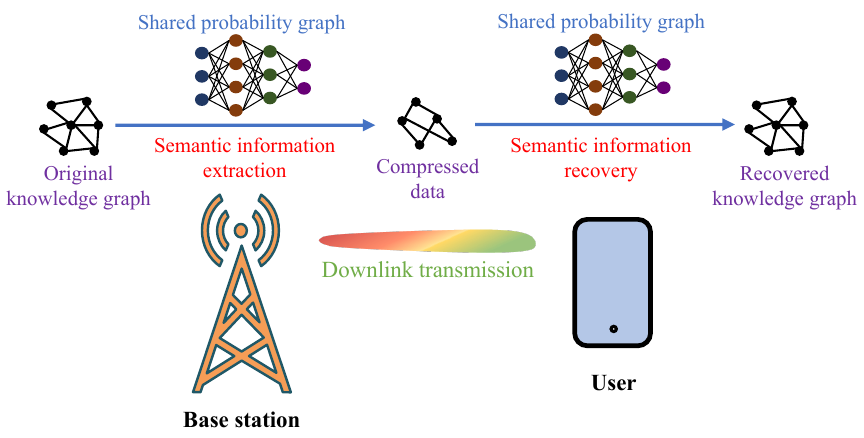}
\caption{Illustration of the considered semantic communication system.}
\label{fig3}
\end{figure}

Consider a semantic wireless communication system with a base station (BS) and a user, as illustrated in Fig. \ref{fig3}. The BS and the user share common background knowledge, and the BS needs to transmit a substantial amount of text data with similar topics to the user while accounting for limited wireless communication resources.

\subsection{The Construction of Probability Graph}
\begin{figure}[t]
\centering
\includegraphics[width=3.4in]{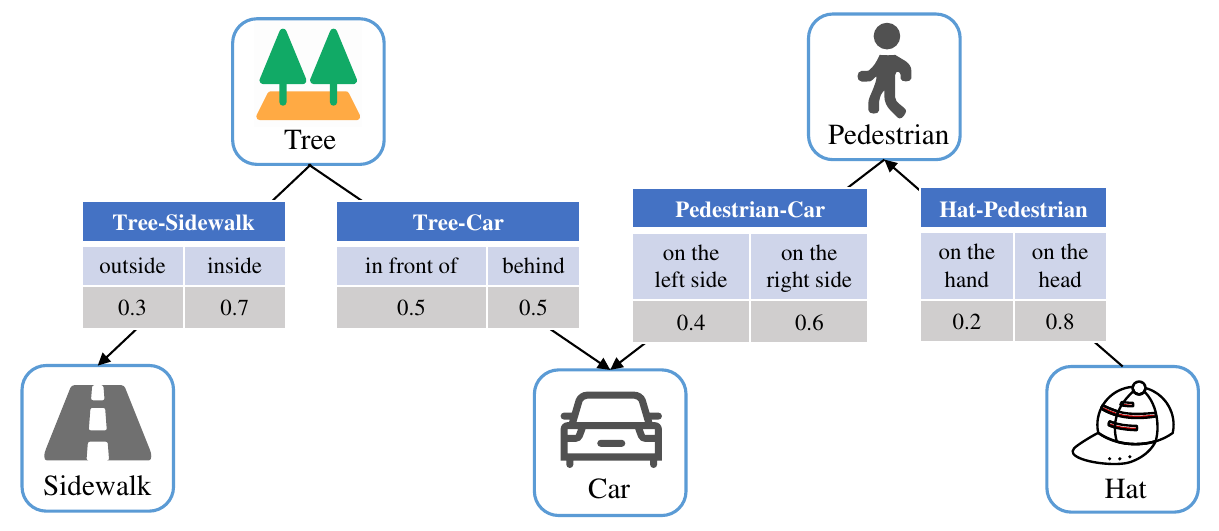}
\caption{Illustration of the probability graph considered in the semantic communication system.}
\label{fig4}
\end{figure}

The semantic information of a knowledge graph is typically expressed as triples in the form of (\emph{head, relation, tail}). From a piece of text data, multiple triples can be extracted, and these triples can be used to characterize a knowledge graph. In cases where there are multiple texts with similar topics, some of the corresponding triples may share the same head and tail entities, but differ in the specific relations.
When there is a lot of text data with similar topics, their corresponding knowledge graphs can be merged into a large probability graph using a statistical method. In this case, the edges in the knowledge graph are not uniquely determined relational edges, but probabilistic edges that combine the frequency of occurrence of different relations, as illustrated in Fig. \ref{fig4}.

Before the formal communication begins, the BS constructs a probability graph based on a large number of samples. Once the construction is complete, the BS sends the probability graph to the user as common background knowledge. This process is performed only once, and the formal communication begins after its completion. During the formal communication, the BS transmits the text data required by the user.

The sample data can be written in the following form:

\vspace{-.5em}
\begin{equation}\label{eq1}\vspace{-.5em}
    \mathcal T = \left\{T_1,T_2, \cdots ,T_n, \cdots ,T_N\right\},
\end{equation}
where $N$ is the total number of sample data, and $\mathcal T$ represents the sample data set. 

The knowledge graph extracted from each sample data $T_n$ is represented as

\vspace{-.5em}
\begin{equation}\label{eq2}\vspace{-.5em}
    G_n = \left\{\varepsilon^1_n,\varepsilon^2_n,\cdots,\varepsilon^m_n,\cdots,\varepsilon^M_n\right\},
\end{equation}
where $\varepsilon^m_n$ is the $m$th triple in knowledge graph $G_n$, and $M$ is the total number of triples. The triple $\varepsilon^m_n$ can be written in the following form:

\vspace{-.5em}
\begin{equation}\label{eq3}\vspace{-.5em}
    \varepsilon^m_n = \left(h^m_n, r^m_n, t^m_n\right),
\end{equation}
where $h^m_n$ is the head entity of triple $\varepsilon^m_n$, $t^m_n$ is the tail entity of triple $\varepsilon^m_n$, and $r^m_n$ is the relation of $h^m_n$ and $t^m_n$.

The probability graph shared by the BS and the user can be represented as

\vspace{-.5em}
\begin{equation}\label{eq4}\vspace{-.5em}
    \mathbb{G} = \left\{\delta_1, \delta_2, \cdots, \delta_s, \cdots, \delta_S\right\},
\end{equation}
where $\delta_s$ is the quadruple with relational probability, and $S$ is the total number of quadruples. Specifically, $\delta_s$ can be written in the following form:

\vspace{-1em}
\begin{equation}\label{eq5}\vspace{-.5em}
    \delta_s=\left\{h_s,\left[\left(r_s^1, \mathcal{N}_s^1\right), \cdots,\left(r_s^i, \mathcal{N}_s^i\right), \cdots,\left(r_s^I, \mathcal{N}_s^I\right)\right], t_s\right\},
\vspace{.5em}
\end{equation}
where $h_s$ is the head entity of quadruple $\delta_s$, $t_s$ is the tail entity of quadruple $\delta_s$, $r_s^i$ is the $i$th relation that exists between $h_s$ and $t_s$, $I$ is the total number of different relations that exists between $h_s$ and $t_s$, and $\mathcal{N}_s^i$ is the set of samples in which the triple $\left(h_s, r^i_s, t_s\right)$ holds. For example, if $\left(h_s, r^i_s, t_s\right)$ holds in sample $T_1$, $T_4$, and $T_7$, then $\mathcal{N}_s^i = \left\{1,4,7\right\}$.

By now, the construction of the probability graph shared by the BS and the user is completed.

\subsection{Probability Graph-enabled Information Compression}
Once the set of quadruples has been obtained, it is possible to calculate the multidimensional conditional probability distribution in the presence of multiple sources of information. The methodology will be elaborated upon in the following.

If no prior information is available, the probability that the triple $\left(h_s, r^i_s, t_s\right)$ holds can be written as

\vspace{-.5em}
\begin{equation}\label{eq6}\vspace{-.5em}
    p\left(h_s, r_s^i, t_s\right)=\frac{\operatorname{num}\left(\mathcal{N}_s^i\right)}{\sum_{i=1}^I \operatorname{num}\left(\mathcal{N}_s^i\right)},
\end{equation}
where $\operatorname{num}\left(\mathcal{N}_s^i\right)$ is the number of elements in the set $\mathcal{N}_s^i$.
If the triple $\left(h_s, r^i_s, t_s\right)$ is known to be true, then the conditional probability of the triple $\left(h_k, r^i_k, t_k\right)$ can be expressed as follows:

\vspace{-1em}
\begin{equation}\label{eq7}\vspace{-.5em}
    p\left[\left(h_k, r_k^i, t_k\right) \mid\left(h_s, r_s^i, t_s\right)\right]=\frac{\operatorname{num}\left(\mathcal{N}_k^i \cap \mathcal{N}_s^i\right)}{\operatorname{num}\left[\mathcal{N}_s^i \cap\left(\bigcup_{i=1}^I \mathcal{N}_k^i\right)\right]},
\vspace{.5em}
\end{equation}
where $\mathcal{N}_k^i \cap \mathcal{N}_s^i$ is the intersection of set $\mathcal{N}_k^i$ and set $\mathcal{N}_s^i$, and $\bigcup_{i=1}^I \mathcal{N}_k^i$ is the union of set $\mathcal{N}_k^1$ to $\mathcal{N}_k^I$.

Using a similar approach, we can derive multidimensional conditional probability.


When transmitting the knowledge graph that corresponds to a particular text data, the multidimensional conditional probability distribution described above can be used to remove semantic relations for the purpose of information compression. The compression process can be achieved using the following scheme.

Prior to transmitting the text data, the BS first applies natural language processing techniques to extract the corresponding knowledge graph denoted as $G$, which is expressed as follows:

\vspace{-.5em}
\begin{equation}\label{eq9}\vspace{-.5em}
    G = \left\{\varepsilon_1, \varepsilon_2, \cdots, \varepsilon_j, \cdots, \varepsilon_J\right\},
\end{equation}
where $\varepsilon_j$ is the $j$th triple in $G$, and $J$ is the total number of triples in $G$. Specifically, the triple $\varepsilon_j$ can be written in the following form:
%
\begin{equation}\label{eq10}\vspace{-.5em}
    \varepsilon_j = \left(h_j,r_j,t_j\right).
\end{equation}
Next, the knowledge graph $G$ is compared with the shared probability graph $\mathbb{G}$, and the comparison process can be divided into several rounds.

In the initial round, there is no prior information. For any $\varepsilon_j$, we need to examine if there exists $\delta_s$ in $\mathbb{G}$ where $h_j=h_s$ and $t_j=t_s$. If such a case is found, we proceed to investigate whether there exists a relation $r_s^i$ within the relation set of $\delta_s$ that matches $r_j$. If a match is identified, we further determine whether the corresponding relation probability $p\left(h_s,r_s^i,t_s \right)$ is the highest among the set of relation probabilities denoted as $P_s$. It is possible to represent $P_s$ as follows:

\vspace{-1em}
\begin{equation}\label{eq11}\vspace{-.5em}
    P_s = \left\{p\left(h_s,r^1_s,t_s\right),\cdots,p\left(h_s,r^i_s,t_s\right),\cdots,p\left(h_s,r^I_s,t_s\right)\right\}.
\vspace{.5em}
\end{equation}
If $p\left(h_s,r_s^i,t_s \right)$ happens to be the largest among the relation probabilities in set $P_s$, we can exclude the relation $r_j$ from the triple $\varepsilon_j=\left( h_j,r_j,t_j \right)$ during transmission. In this case, only the head entity and tail entity need to be sent, reducing the transmitted data volume. However, if there is no $\delta_s$ in $\mathbb{G}$ satisfying $h_j=h_s$ and $t_j=t_s$, or if such a $\delta_s$ exists but there is no relation $r_s^i$ in its relation set that matches $r_j$, then the entire triple must be transmitted without omission.

In the second round, one-dimensional conditional probabilities can be employed as prior information. As certain triples had their semantic relations omitted in the first round, these triples can serve as conditions for the second round's conditional probability search. The second round of comparison follows a similar process as the first round. We can denote the set of triples that were not omitted in the first round and do not necessarily require full transmission as
%
\begin{equation}\label{eq12}\vspace{-.5em}
    G^1 = \left\{\varepsilon_1,\varepsilon_2,\cdots,\varepsilon_a,\cdots,\varepsilon_A\right\}.
\end{equation}
Write the set of triples whose relations were omitted in the first round as
%
\begin{equation}\label{eq13}\vspace{-.5em}
    O^1 = \left\{\epsilon_1,\epsilon_2,\cdots,\epsilon_b,\cdots,\epsilon_B\right\}.
\end{equation}

For every $\varepsilon_a$ in $G^1$, we need to compute its conditional probability given each triple $\epsilon_b$ in $O^1$. Let's assume that $\varepsilon_a$ corresponds to $\delta_a = \left(h_a,r_a^i,t_a \right)$ in the shared probability graph $\mathbb{G}$. The conditional probability can be represented as $p\left[\left(h_a,r_a^i,t_a \right)\mid \epsilon_b \right]$, which we abbreviate as $p^1_{a\mid b}$. Consequently, we can construct a probability matrix consisting of one-dimensional conditional probabilities for $\left(h_a,r_a^i,t_a \right)$ as follows:
%
\vspace{-.5em}
\begin{equation}\label{eq14}\vspace{-.5em}
    P^1_{\delta_a} = \left\{p^1_{i \mid b}\right\}^{B\times I},
\end{equation}
where $I$ is the total number of different relations between $h_a$ and $t_a$ in $\mathbb{G}$, and $B$ is the total number of elements in $O^1$.
In the probability matrix $P^1_{\delta_a}$, we begin with the first row and examine whether $p^1_{i \mid b}$ is the largest element within that row. If it is indeed the largest, we take note of the column number and exclude the relation $r^i_a$ when transmitting the corresponding triple. However, if $p^1_{i \mid b}$ is not the largest element in any of the rows, it indicates that the corresponding relation cannot be omitted during transmission.

We repeat the aforementioned procedure for every triple in $G^1$, aiming to identify the triples that can be omitted. This constitutes the first cycle in the second round. In the second cycle, there will be fewer remaining triples that have not been omitted. We create a new $G^1$ using these unomitted triples and form a new $O^1$ for the newly omitted triples. We continue this process until no new triples can be omitted, iterating through multiple cycles if necessary.

In the third round, we incorporate two-dimensional conditional probabilities as prior information. We derive the corresponding $G^2$ and $O^2$ based on the previous round's results. Subsequently, we calculate a three-dimensional probability matrix, denoted as $P^2$, which comprises two-dimensional conditional probabilities.
This iterative process continues for subsequent rounds, where each round utilizes conditional probabilities with one additional dimension.

The comparison process remains largely unchanged for each round and does not vary significantly. Therefore, it is not reiterated here. However, it is important to note that as the rounds progress, deeper comparisons require increased computational resources. Furthermore, conducting additional rounds of comparison does not necessarily guarantee proportionate benefits. Thus, the number of rounds to be compared can be determined based on the specific communication needs and available computational resources.

Once the semantic information compression is performed by the BS, it transmits the conditions of each information compression to the user, along with the compressed triples. These compression conditions are represented as very short bit streams, which are insignificant compared to the size of the original data to be transmitted. Upon receiving the compressed information, the user can reconstruct the omitted information by leveraging the compressed conditions in conjunction with the common background knowledge $\mathbb{G}$.

\subsection{Joint Communication and Computation Resource Allocation}
The above reasoning process enables the exclusion of certain portions of the knowledge graph from transmission, thereby reducing communication delays. However, compressing information through reasoning requires additional time and energy for computation. To achieve improved results and minimize the overall system energy consumption during the communication process, it is crucial to consider the joint allocation of communication and computation resources. By optimizing the allocation of these resources in a coordinated manner, it is possible to strike a balance that maximizes the efficiency of the system while minimizing energy consumption.

The total latency of the semantic communication system comprises two main components: the communication latency, denoted as $t_1$, and the computation latency, denoted as $t_2$. It is important to note that the latency for processing and transmitting a text data message is constrained by a limit, denoted as $T$. Therefore, in order to meet the latency limit, the total process must adhere to the condition that the sum of communication latency and computation latency is less than or equal to $T$, i.e., $t_1 + t_2 \leq T$. This ensures that the overall latency of the system remains within the specified limit.

The communication time delay $t_1$ is calculated as follows.

Assuming a stable channel between the BS and the user during the communication process, with a transmission bandwidth of $B$, a transmit power of $p$ from the BS, a path loss of $h$, and a noise variance of $\sigma^2$, the channel capacity between the BS and the user can be expressed as follows:
%
\begin{equation}\label{eq15}\vspace{-.5em}
    c = B\log_2\left(1+\frac{ph}{\sigma^2}\right).
\end{equation}

If the head entity, relation, and tail entity in each triple are all encoded with the same number of bits $R$, then the total number of bits of semantic information $D$ in the text data can be calculated as
%
\begin{equation}\label{eq16}\vspace{-.5em}
    \operatorname{size}\left(D\right) = R\left(3M-E\right),
\end{equation}
where $M$ is the total number of triples in $D$, and $E$ is the number of triples in which the relation is omitted.
Consequently, the communication latency can be written as
%
\begin{equation}\label{eq17}\vspace{-.5em}
    t_1 = \frac{\operatorname{size}\left(D\right)}{c} = \frac{R\left(3M-E\right)}{B\log_2\left(1+\frac{ph}{\sigma^2}\right)}.
\end{equation}


The computational load of the proposed knowledge graph semantic information compression system is directly related to the number of comparisons in the compression process. To simplify the analysis, we will only consider the first and second rounds of comparison as described in the previous subsection. As it is not feasible to pre-determine the computational resources required to omit $E$ relations for a specific text data, we employ a statistical approach to calculate the ratio of the number of triples that can be omitted in each round to the total number of remaining triples. Based on this ratio, we estimate the computational resources required to omit $E$ relations in a statistical sense.

The ratio of the number of triples that can be omitted more in each round to the total number of triples before this round can be expressed as

\vspace{-.5em}
\begin{equation}\label{eq18}\vspace{-.5em}
    Q = \left\{q_1,q_2,\cdots,q_n,\cdots,q_N\right\},
\end{equation}
where $q_1$ is the ratio of the number of triples that can be omitted more in the first round to the total number of all triples, $q_2$ is the ratio of the number of triples that can be omitted more in the first cycle of the second round to the number of triples whose relation is not omitted, and $q_n$ is the ratio of the number of triples that can be omitted more in the $\left(n-1\right)$th cycle of the second round to the number of triples whose relation is not omitted before this cycle.

We can express the number of semantic relations that can be omitted in each round or cycle using a recursive formula:

\vspace{-.5em}
\begin{equation}\label{eq19}\vspace{-.5em}
    \left\{\begin{array}{l}
        E_1=M q_1, \\
        E_2=\left(M-E_1\right) q_2, \\
        E_3=\left(M-E_1-E_2\right) q_3, \\
        \vdots \\
        E_N=\left(M-E_1-E_2-\cdots-E_{N-1}\right) q_N,
    \end{array}\right.,
\end{equation}
where $M$ is the total number of triples, and $E_N$ is the number of relations which can be omitted in the $\left(N-1\right)$th cycle of the second round.

Based on (\ref{eq19}), the computation load to omit $E$ relations in $M$ triples can be written as

\vspace{-.5em}
\begin{equation}\label{eq20}\vspace{-.5em}
    l\left(E\right)=\left\{\begin{array}{l}
        \frac{E}{q_1}, 0 \leq E \leq E_1, \\
        \frac{E_1}{q_1}+\frac{\left(E-E_1\right) E_1}{q_2}, E_1<E \leq E_1+E_2, \\
        \frac{E_1}{q_1}+\left(\frac{E_1}{q_1}-E_1\right) E_1+\\
        \frac{\left(E-E_2-E_1\right) E_2}{q_3}, E_1+E_2<E \leq E_1+E_2+E_3, \\
        \vdots \\
        \frac{E_1}{q_1}+\left(\frac{E_1}{q_1}-E_1\right) E_1
        \\+\left(\frac{E_1}{q_1}-E_1-E_2\right) E_2\\+\cdots+\frac{\left(E-E_{N-1}-\cdots-E_1\right) E_{N-1}}{q_N}, \\\sum_{n=1}^{N-1} E_n<E \leq \sum_{n=1}^N E_n.
    \end{array}\right..
\vspace{.5em}
\end{equation}
It can be observed that $l\left(E\right)$ is an increasing segmented function, where each segment is a linear function with respect to $E$.
Consequently, the computation latency can be written as
%
\begin{equation}\label{eq21}\vspace{-.5em}
    t_2 = \frac{\tau_1 l\left(E\right)}{f},
\end{equation}
where $\tau_1$ is a constant coefficient, and $f$ is the computing capacity.

\section{Problem Formulation}
The communication energy consumption can be written as
%
\begin{equation}\label{eq22}\vspace{-.5em}
    e_1 = t_1 p.
\end{equation}
The computation energy consumption can be written as\cite{10032275}
%
\begin{equation}\label{eq23}\vspace{-.5em}
    e_2 = \tau_1 \tau_2 l\left(E\right) f^2,
\end{equation}
where $\tau_2$ is a constant coefficient.

Based on the considered model, we can construct the following joint optimization problem:
%
\begin{subequations}\label{eq24}\vspace{-.5em}
    \begin{align}
        \min_{p, E} \quad & e_1+e_2, \tag{\ref{eq24}}\\
        \textrm{s.t.} \quad & \frac{R(3 M-E)}{B \log _2\left(1+\frac{p h}{\sigma^2}\right)}+\frac{\tau_1 l\left(E\right)}{f} \leq T, \label{c1}\\
        & 0 \leq p \leq p_{\text{max}}, \label{c2}\\
        & 0 \leq E \leq M, E \in \mathbb{N} \label{c3}.
    \end{align}
\end{subequations}

The optimization problem comprises of three constraints. Constraint (\ref{c1}) limits the total delay of communication and computation, while constraint (\ref{c2}) limits the transmit power $p$ of the BS to a non-negative value that does not exceed a maximum limit $p_{\text{max}}$. Constraint (\ref{c3}) limits the total number of omitted relations $E$ to a non-negative integer that does not exceed the total number of triples $M$. The objective of the optimization problem is to minimize the total energy consumption of the semantic communication system.
Problem \eqref{eq24} is nonconvex, which can be solved via exhaustive searching the discrete variable $E$ and gradient descent optimization method to optimize $p$ with each give $E$.

\section{Simulation Results and Analysis}
In the simulations, the maximum transmit power of the BS is 30 dBm. Unless specified otherwise, we set the bandwidth of the BS $B=10$ MHz, total number of triples $M=100$, and total latency limit $T=1$ ms. The values of most of the simulation parameters are set based on \cite{10032275}. 

To evaluate the performance of the proposed joint communication and computation optimization system based on probability graph, labeled as `JCCPG', we designed two baseline algorithms for comparison. `Traditional' involves directly transmitting all triples with the same total delay, without considering the maximum transmission power limit. `Simplified JCCPG' optimizes the allocation of communication and computation resources based solely on the results of the first round of comparison.

\begin{figure}[tb]
\centering
\includegraphics[width=2.8in]{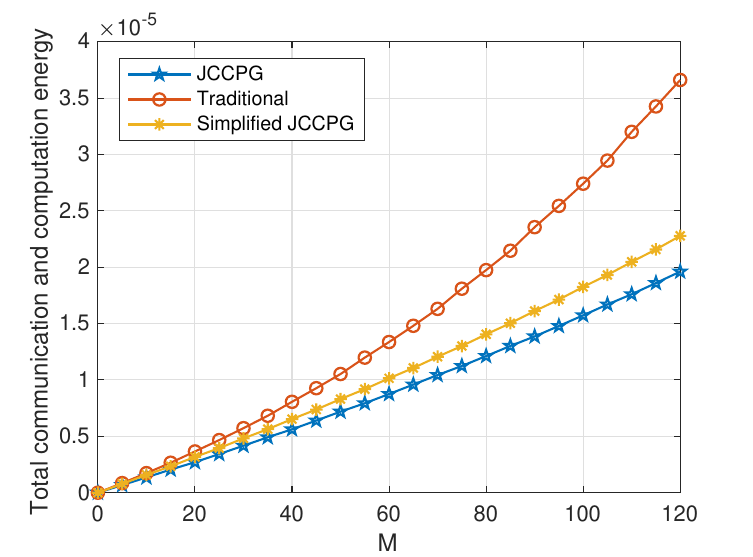}
\caption{Total communication and computation energy vs. total number of transmitted triples.}
\label{fig6}
\end{figure}

Fig. \ref{fig6} illustrates the variation trend of the total communication and computation energy consumption in the system with respect to the total number of transmitted triples. As the total number of transmitted triples increases, the energy consumption of all algorithms in the system also increases due to the increased transmission and computation requirements. However, the proposed algorithm and `Simplified JCCPG' exhibit a significantly smaller growth rate compared to `Traditional'. This indicates the robustness of the proposed algorithm. Furthermore, the total system energy consumption of the proposed algorithm is consistently lower than that of both `Traditional' and `Simplified JCCPG'. This advantage becomes more pronounced as the total number of transmitted triples increases. These findings highlight the significant advantage of the proposed algorithm in transmitting the large-scale data.

\section{Conclusion}
In this paper, we have proposed a novel probability graph-based semantic information compression system designed for scenarios where the BS and the user share common background knowledge. By leveraging probability graph, we established a framework to represent and compress shared knowledge between the communicating parties. 
In conclusion, our work presented a comprehensive framework for probability graph-based semantic information compression, addressing joint communication and computation resource optimization challenge in semantic communication systems.

\bibliographystyle{IEEEtran}
\bibliography{IEEEabrv,conference}

\begin{thebibliography}{10}
\providecommand{\url}[1]{#1}
\csname url@samestyle\endcsname
\providecommand{\newblock}{\relax}
\providecommand{\bibinfo}[2]{#2}
\providecommand{\BIBentrySTDinterwordspacing}{\spaceskip=0pt\relax}
\providecommand{\BIBentryALTinterwordstretchfactor}{4}
\providecommand{\BIBentryALTinterwordspacing}{\spaceskip=\fontdimen2\font plus
\BIBentryALTinterwordstretchfactor\fontdimen3\font minus \fontdimen4\font\relax}
\providecommand{\BIBforeignlanguage}[2]{{%
\expandafter\ifx\csname l@#1\endcsname\relax
\typeout{** WARNING: IEEEtran.bst: No hyphenation pattern has been}%
\typeout{** loaded for the language `#1'. Using the pattern for}%
\typeout{** the default language instead.}%
\else
\language=\csname l@#1\endcsname
\fi
#2}}
\providecommand{\BIBdecl}{\relax}
\BIBdecl

\bibitem{larsson2014massive}
E.~G. Larsson, O.~Edfors, F.~Tufvesson, and T.~L. Marzetta, ``Massive mimo for next generation wireless systems,'' \emph{IEEE communications magazine}, vol.~52, no.~2, pp. 186--195, 2014.

\bibitem{2023big}
Z.~Chen, Z.~Zhang, and Z.~Yang, ``Big {AI} models for 6g wireless networks: Opportunities, challenges, and research directions,'' \emph{arXiv preprint arXiv:2308.06250}, 2023.

\bibitem{gunduz2022beyond}
D.~G{\"u}nd{\"u}z, Z.~Qin, I.~E. Aguerri, H.~S. Dhillon, Z.~Yang, A.~Yener, K.~K. Wong, and C.-B. Chae, ``Beyond transmitting bits: Context, semantics, and task-oriented communications,'' \emph{IEEE Journal on Selected Areas in Communications}, vol.~41, no.~1, pp. 5--41, 2022.

\bibitem{10024766}
W.~Xu, Z.~Yang, D.~W.~K. Ng, M.~Levorato, Y.~C. Eldar, and M.~Debbah, ``Edge learning for {B5G} networks with distributed signal processing: Semantic communication, edge computing, and wireless sensing,'' \emph{IEEE J. Sel. Topics Signal Process.}, vol.~17, no.~1, pp. 9--39, Jan. 2023.

\bibitem{chaccour2022less}
C.~Chaccour, W.~Saad, M.~Debbah, Z.~Han, and H.~V. Poor, ``Less data, more knowledge: Building next generation semantic communication networks,'' \emph{arXiv preprint arXiv:2211.14343}, 2022.

\bibitem{9416312}
S.~Ji, S.~Pan, E.~Cambria, P.~Marttinen, and P.~S. Yu, ``A survey on knowledge graphs: Representation, acquisition, and applications,'' \emph{IEEE Transactions on Neural Networks and Learning Systems}, vol.~33, no.~2, pp. 494--514, 2022.

\bibitem{xie2021deep}
H.~Xie, Z.~Qin, G.~Y. Li, and B.-H. Juang, ``Deep learning enabled semantic communication systems,'' \emph{IEEE Transactions on Signal Processing}, vol.~69, pp. 2663--2675, 2021.

\bibitem{weng2021semantic}
Z.~Weng and Z.~Qin, ``Semantic communication systems for speech transmission,'' \emph{IEEE Journal on Selected Areas in Communications}, vol.~39, no.~8, pp. 2434--2444, 2021.

\bibitem{tong2021federated}
H.~Tong, Z.~Yang, S.~Wang, Y.~Hu, W.~Saad, and C.~Yin, ``Federated learning based audio semantic communication over wireless networks,'' in \emph{2021 IEEE Global Communications Conference (GLOBECOM)}.\hskip 1em plus 0.5em minus 0.4em\relax IEEE, 2021, pp. 1--6.

\bibitem{10032275}
Z.~Yang, M.~Chen, Z.~Zhang, and C.~Huang, ``Energy efficient semantic communication over wireless networks with rate splitting,'' \emph{IEEE Journal on Selected Areas in Communications}, vol.~41, no.~5, pp. 1484--1495, 2023.

\end{thebibliography}

\end{document}